\documentstyle [12pt] {article}

\begin{document}
\begin{titlepage}
\vskip 2cm
\title { Nilpotent Gauging of SL(2,R)$WZNW$ Models,
\\ and Liouville Field }
\author { M.Alimohammadi $^a$ , F.Ardalan $^{a,b}$ , H.Arfaei $^{b,c}$ }
\date{}
\maketitle
\begin {center}
$^a$ {\it Institute for studies in Theoretical Physics and Mathematics} \\
 {\it P.O.Box 19395-1795, Tehran, Iran} \\
$^b$ {\it Physics Department, Sharif University of Technology} \\
{\it P.O.Box 11365-9161, Tehran, Iran}\\
$^c$ {\it  Physics Institute, Bonn University}\\
{\it Nussalle 12, 5300 Bonn 1 , Germany}
\end{center}
\vskip 1cm

\begin{abstract}
{We consider the gauging of $ SL(2,R) $ WZNW model by its
nilpotent subgroup E(1). The resulting space-time of the corresponding
sigma model
 is seen to collapse to a one dimensional field theory of Liouville.
Gauging the diagonal  subgroup $ E(1) \times U(1) $ of $SL(2,R)
\times U(1)$ theory yields an extremal three
dimensional black string. We show that these solutions are obtained from
the two  dimensional black hole
of Witten and the three dimensional black string of Horne and Horowitz
by boosting the gauge group.}
\end{abstract}
\vskip 2cm
\vfill

\ Physikalisches Institut  \hfill {hep-th/9304024  \ \ \ \ \ \ \ }

\ Nussalle 12 \hfill  {BONN-HE-93-12 \ \ \ \ \ \ \ }

\ 5300 Bonn 1  \hfill {SUTDP-93/72/3 \ \ \ \ \ \ \ }

\ Germany   \hfill    {IPM-93-007 \ \ \ \ \ \ \ }

\hfill                {March 1993 \ \ \ \ \ \ \ }
\end{titlepage}

\noindent
\section{Introduction}

There has been a great deal of interest in the gauged
Wess-Zumino-Novikov-Witten
models as candidates for description of non trivial geometric back grounds
for string theory, since the discovery by Witten [1] that the target
manifold of the conformal field theory of the gauged $SL(2,R) \ WZNW$ model
was that of a two dimensional black hole [2-12].
The original model which had a singularity and a horizon was obtained by
gauging a non-compact
$U(1)$ subgroup of $SL(2,R)$  whereas gauging a compact subgroup would
yields a non singular manifold.

The extension of the construction to higher dimensions produced charged
black holes[3-10]. In particular Horne and Horowitz [4], considered the
simplest extension to $SL(2,R)$ \\ $\times U(1)$ and
constructed various three dimensional black strings, carrying charge.
The duality relation between to two types of gauging, the vector and
 axial gauging has also been extensively studied [2,3,12].

Gauging two conjugate subgroups of a $WZNW$ model yields identical
theories.
But guaging it with two subgroups of different conjugacy class should
yield different target spaces. Thus
the Lorentzian two dimensional black hole solution of Witten [1], with
singularity and horizon, obtained from gauging the non-compact subgroup
 of $SL(2,R)$ , is
distinct from the Euclidean black hole target manifold obtained by
gauging a
compact subgroup, which is in a different conjugacy class of the subgroups
of $SL(2,R)$.
Therefore it is interesting to consider gauging by an element of the
only other
conjugacy class of $SL(2,R)$, the class of nilpotent subgroup $E(1)$.

In this paper \ we undertake to study the target manifold of the $SL(2,R)$
and $SL(2,R)$ \\ $\times U(1) \ WZNW$ models gauged by their $E(1)$
subgroups. In the first case we find
an unexpected result , that is the resultant manifold is only one
dimensional. In fact the effective action turns out to be the Liouville
field action.
 To understand the result we study the Witten black hole with the gauged
 subgroup boosted. In the limit of infinite
boost we obtain the Liouville field as expected from gauging the
$E(1)$ subgroup. We will also discuss the connection of our result
with the standard Hamiltonian reduction [13-15].

 Next we consider the gauging of the $SL(2,R)\times U(1) \ WZNW $ model
 where the gauge group is the diagonal $ U(1)\times E(1)$ subgroup .
In this case  no such  reduction of the degrees of freedom occurs ;
instead, we obtain the extremal
 three dimensional black string solution which is again found  by an
 infinite boost of  the noncompact gauging.

In the $SL(2,R)$ case, for large but finite boost parameter, the theory
 remains two dimensional and may be interpreted as a $c=1$ matter
coupled to Liouville, which acts as a slowly varying back ground [16].

In section 2 we discuss all the possible gaugings of $SL(2,R)$ and
boosting of the black hole solutions. In section 3 the $E(1)$ gauging
of the $SL(2,R)\times U(1) \ WZNW$ model is considered
and its singularity structure discussed. In section 4 we will make
additional comments about the
symmetries of the $SL(2,R)  / E(1)$ theory and discuss possible reasons for
its reduction
of degrees of freedom.

\section{Different Gaugings of $SL (2,R)$ $WZNW$ Model}
\setcounter{equation}{0}

In this section we consider gauging the $G=SL(2,R)$ Wess-Zumino-Witten
models, by one of its subgroups $H$.  $SL(2,R)$ has
three distinct conjugacy classes of subgroups isomorphic to: rotations
in two dimension $SO(2)$, Lorentz transformations in two dimension
$SO(1,1)$, and the isotropy group of light-like vectors , i.e. the one
dimensional eucledian group $E(1)$.  In four dimensional Lorentz
group, the corresponding three subgroups are
$SO(3)$ , $SO(2,1)$ and $E(2)$, the last of which is the group of motions
of the two dimensional plane and is the symmetry of a massless particle.

We take $\sigma _1, i\sigma _2, $ and $\sigma _3$, where $\sigma _i$ are
the Pauli matrices, as the generators of $SL(2,R)$. Then the subgroup
$SO(2)$ is
generated by $i \sigma _2$, the $SO(1,1)$ by $\sigma _3$, and $E(1)$
is generated by $\sigma ^+ = \sigma _3 + i \sigma _2$. Observe that
$\sigma^{+2} =0$ and thus $E(1)$ is nilpotent.

In  a $WZNW$ model with the group $G$,  two different ways of
gauging an abelian subgroup $H$ has been considered : axial and vector
gauging. For the axial gauging the group action  on the $SL(2,R)$ is
\begin{equation}
g \longrightarrow hgh \ \ , \ \ \ \ g\epsilon G  \ \ , \ h \epsilon H.
\end{equation}
Then the action,
\begin{equation}
\begin{array}{ll}
I(g,{\bf A)} & {\displaystyle =I_{WZNW} + \frac {k} {2 \pi } \int d^2 z
tr (\overline {\bf A} g^{-1} \partial g +
{\bf A}\overline { \partial } g g^{-1} +  {\bf A}\overline {\bf A}+ g^{-1}
{\bf A}g \overline {\bf A} ),} \\
I_{WZNW} & {\displaystyle = \frac {k}{4 \pi} \int _ \Sigma d^2z
tr (g^{-1} \partial g g^{-1} \overline {\partial }g )
-\frac {k}{12 \pi} \int _B d^3x tr (g^{-1} dg \wedge g^{-1} dg
\wedge g^{-1}
dg)}\\
\end{array}
\end{equation}
is invariant under (2.1) together with,
\begin{equation}
\begin{array}{ll}
 {\bf  A}\longrightarrow & h({\bf  A}+ \partial ) h^{-1}\\
\overline {\bf A} \longrightarrow & h^{-1} ( {\overline {\bf  A} }+
\overline {\partial }) h,\\
\end{array}
\end{equation}
where $ {\bf A}$ and $\overline  {\bf A}$ are the complex components
of the gauge field  and take their
values in the Lie algebra of $H$. We will  denote
the coefficient of the Lie algebra generator by $A$.

Similarly, the action
\begin{equation}
I(g,{\bf A})=I_{WZNW} + \frac {k}{2 \pi} \int d^2 z tr ({\bf A}
\overline {\partial } g g ^{-1} -
\overline {{\bf A}} g^{-1} \partial g + {\bf A}\overline {{\bf A}} -
g ^{-1} {\bf A}g \overline {{\bf A}})
\end{equation}
is invariant under the vector gauge transformation ,
\begin{equation}
\begin{array}{ll}
g \longrightarrow & hgh^{-1}\\
{\bf A} \longrightarrow & h^{-1}({\bf A}+ \partial ) h  \\
\overline {{\bf A}} \longrightarrow & h^{-1}( \overline {{\bf A}} +
\overline {\partial } )h
\end{array}
\end{equation}

In the following we will first consider the axial gauging (1-3) for
$G=SL(2,R)$ and its three different subgroups. Then we will take up the
corresponding
vector gaugings, and at last we will discuss the duality between the two
gaugings.

Let us take the parametrization
\begin{equation}
g= \left ( \begin {array} {ll}
a & u \\
-v & b
\end {array} \right ) \ \ \ ab + uv=1
\end{equation}
for the group element $g \epsilon SL(2,R)$ and gauge the action (2) with
the noncompact subgroup $SO(1,1)$ generated by $\sigma _3$. Using the gauge
freedom  we can take
\begin{equation}
a+b=0
\end{equation}
when $ab<0 $ , and
\begin{equation}
a-b=0
\end{equation}
when $ab>0$. Note that $u$ and $v$ are gauge invariant parameters.
Substituting in (2.2), then
yields
\begin{equation}
I(g,A)= \frac {2k}{\pi } \int d^2 z (1-uv) [A- \frac {u \partial v - v
\partial u }{ 4(1-uv)} ]
[ \overline {A} + \frac {u \overline {\partial} v - v \overline
{\partial } u}{ 4 (1-uv)} ]
-\frac {k}{4 \pi} \int d^2 z \frac { \partial u \overline {\partial } v -
\partial v \overline {\partial } u}{ 1-uv}
\end{equation}
The second term, when compared to a target space model, gives the two
dimensional metric
\begin{equation}
ds^2 = -\frac {k}{2} \ \ \frac {du dv}{1-uv}
\end{equation}
The first term, after integration over the gauge field, gives the
dilaton field
\begin{equation}
\Phi = ln (1-uv) +a
\end{equation}
wher $a$ is a constant The above metric and dilaton fields can also be
verified to satisfy the  the equations motion of the
target space action.

We have used the following target space action which we set down for
future
reference,
\begin{equation}
I= -\frac {1}{4 \pi \alpha '} \int d ^2 \sigma \sqrt {h} (h ^{\alpha
\beta} g _{\mu \nu} + \epsilon ^{\alpha \beta } B_{\mu \nu})
\partial _\alpha X^\mu \partial _\beta X^\nu
+\frac {1}{4 \pi } \int d ^2 \sigma \sqrt {h} (\frac {1}{2} h ^{\alpha
\beta}
\partial _\alpha \Phi \partial _\beta \Phi + \Phi R^{(2)})
\end{equation}
The equation to solve for the dilaton is
\begin{equation}
R_{\mu \nu} =D_{\mu} D_{\nu} \Phi
\end{equation}
The solution (2.10) has been discussed in great detail and corresponds
to a
two dimensional black hole with the singularity at $uv=1$, and horizon
at
$u=0$ and $v=0$. (Fig.1). As seen from Eq.(2.11), the string coupling
vanishes
at spatial infinity and blows up near the singularity. The black hole's
mass is
obtained by ADM procedure and is
\begin{equation}
M= \sqrt {\frac {2}{k}} e^a
\end{equation}

Next we consider the compact subgroup $SO(2)$ generated by $i \sigma _2$.
The
gauge freedom (2.1) allows us to fix
\begin{equation}
u+v=0
\end{equation}
in the notation of (2.6). Then in terms of the gauge invariant
parameters $x=u-v$ and
$y=a-b$, the resultant target space metric and dilaton fields become
\begin{equation}
ds^2 = \ \ \frac {k}{2} \ \ \frac {dx^2 + dy^2}{4+x^2 +y^2}
\end{equation}
\begin{equation}
\Phi = ln (4+x^2 +y^2)+\ {\rm const}.
\end{equation}
This solution is of course nonsingular and is termed Euclidean blackhole.

Now we proceed to the new case of the nilpotent subgruop $E(1)$ of
$SL(2,R)$,
generated by $\sigma = \sigma _3 + i \sigma _2$. The parameters
$x=u-v$ and $w=a-b-u-v$ remain invariant under the axial gauge
transformation
$g \longrightarrow hgh$; and provide a gauge invariant parametrization of
the  quotient manifold $SL(2,R) /E(1)$.  We  fix the gauge condition
\begin{equation}
a+b=0
\end{equation}
which is possible only when $w \neq 0$. However for the region near
$w=0$,
another
gauge fixing is possible:
\begin{equation}
u+v=0
\end{equation}
which describes a one dimensional section of the manifold. Substituting
for various terms in the
action (2.2), taking account of the condition $a+b=0$, we get:
\begin{equation}
I(g,A)=- \frac {k}{2 \pi} \int d^2z \frac {1}{w^2} \left| A+
 \frac {1}{2w^2} (x \partial w - w \partial x) \right |^2
+ \frac {k}{2 \pi} \int d^2 z \frac {|\partial w|^2}{w^2}
\end{equation}
Then integrating over the gauge field A, results in the target space
metric
\begin{equation}
ds^2= \frac {k}{2} \ \ \frac {(dw)^2}{w^2}
\end{equation}
and the dilation field
\begin{equation}
\Phi = 2  ln \ w+a
\end{equation}
where $a$ is a constant. The dilaton form (2.22) can also be obtained from
the differential
equation (13).

The striking feature of our result (21) and (22) is that, both the metric
and the dilaton
field are one-dimensional, that is they depend only on $w$ and not on the
 other
dimension $x$ of the manifold. One of the dimensions of the geometric
manifold $SL(2,R)/ E(1)$ has
evaporated .

Even more interesting, the effctive target model action, where
 $w= \pm e ^\varphi $
depending on the sign of $w$,
\begin{equation}
I_{eff} = \frac {k}{2 \pi} \int d^2 z ( \partial \varphi \overline
{\partial } \varphi + 2 \varphi ),
\end{equation}
is nothing but the action of  a one dimensional bosonic field with
background charge
and  without cosmological
constant
which can be interpreted as Liouville field  in view of the limiting
form of its  coupling to the coordinate $X$ when the boost parameter
considered in the following goes to infinity.
 Although Liouville field has been obtained before , as a
quotient of
$SL(2,R)$ by its subgroups, or as a hamiltonian reduction of the
$SL(2,R) \ WZNW$ model,
 we must emphasize  that the above  result is different  and
unexpected.This is because in these approaches
two degrees of freedom are removed by either a constraint together with a
 gauge
fixing [2,13-15] , or by taking independent left and right group action
$g \longrightarrow h_L gh_R$ for
the gauge transformations [13] where the left and right actions are
effected by independent upper and lower triangular matrices. But here we
do not do anymore than the usual
gauged $WZNW$ models. In the present case the geometric manifold
$SL(2,R)/E(1)$ has two dimensions whereas the string moving on it sees a
 one dimensional manifold.

The reduction in the degrees of freedom in the metric (21) is plausible on
account of the
behaviour of corresponding metrics (10) and (16) for the Lorentzian and
Euclidean
cases, since in appropriate variables [1], there is a change in the
signature  of the metric  when crossing from the
Euclidean  to the Lorentzian metric. However, the complete disappearance
of the
coordinate $x$ from the action (2.23) needs a more exact explanation.

To understand this phenomenon of the disappearance of one of the
 dimensions , and
to understand what has happened to the missing degree of freedom we will
obtain our  solution (2.23) from the standard black hole solution
(10)and (11). We thus boost the subgroup $SO(1,1)$, generated by
$\sigma _3$,
in the direction of $\sigma _1$ by a parameter $t$ . We let t go to
infinity. As boosting the subgroup only shifts it in the same conjugacy
 class,
the effect on the blackhole will simply be an isometric translation of
 the
manifold. However at the limit of the infinite boost, the subgroup
$SO(1,1)$
will degenerate to the nilpotent subgroup $E(1)$ and the black hole
should
correspondingly degenerate into our solution (23), i.e., to the pure
Liouville field. Note that infinite boosting is a singular transformation
and can transform elements of different congugacy classes to each other.

Thus consider ,
\begin{equation}
\begin{array}{ll}
\sigma _3^t & = e^{-\frac {t}{2} \sigma _1} \sigma _3
e ^{\frac {t}{2} \sigma _1}\\
 & = cht \ \sigma _3 + \ sht \ i \sigma _2
\end{array}
\end{equation}
and use the group generated by $\sigma _3 ^t$ to gauge $SL(2,R) \ WZNW$
model,
we get the standard  black hole solution (10) and (11) boosted, which for
large
$t$ gives the following action,
\begin{equation}
I_{eff}= \frac {k}{2\pi } \int d^2 z
 [\overline {\partial} \varphi \partial \varphi +
2 \varphi - 4 \epsilon e ^{-2 \varphi } \overline {\partial }x \partial
x - 4 \epsilon e^{-2 \varphi}
(x^2 +4)(\overline {\partial } \varphi \partial \varphi + 1 )],
\end{equation}
where $\epsilon = e^{-2 t}$ .

 We have chosen the gauge fixing
condition
\begin{equation}
a+b=0
\end{equation}
which is valid for
\begin{equation}
|(a-b)coht- (u+v) sht |>|a+b|
\end{equation}
Otherwise the other gauge fixing condition,
\begin{equation}
(a-b)coht = (u+v)sht
\end{equation}
should be used.

As it can be seen from (2.25), at the limit of $t\longrightarrow \infty
,\ (\epsilon \longrightarrow 0)$, the black hole
effective action approaches the Liouville action (2.23), which we found
as the target
space action of the $SL(2,R) \ \  WZNW$ model gauged by its nilpotent
 subgroup $E(1)$.

If we take the other region of the manifold described by the gauge
condition (2.28), we find an
effective action ,
\begin{equation}
I_{eff} = + \frac {k}{2\pi} \int d^2z \frac {\overline {\partial } x
\partial x}{4+ x^2} .
\end{equation}

Now we may trace the cause of the elimination of the extra degree of
 freedom
$x$, when $t\longrightarrow \infty$: As it was pointed out before, for
 finite $t$ there are
two distinct regions of $(u,v)$ plane determined by the gauge conditions
(2.26)
and (2.28). When $t$ increase, the region described by the gauge condition
(2.26), $a+b=0$, narrows
and tends to the one dimensional line, parametrized by $\varphi $
above. What then happens to the other region determined by the second
gauge condition
(2.27)? The answer is that, exactly at $t= \infty$, $(\sigma $,
nilpotent gauging),
the condition $a+b=0$, suffices to describe all the regions; i.e.,
at $t=\infty $
the gauge orbits of the two region connect to yield a single gauge
orbit of the nilpotent
subgroup E(1), thus rendering the theory one dimensional.

In the effective action (25), the nonleading terms have the form of
the interactions of
a $\ c=1$ matter field $x$ with the Liovuille field $\varphi$. In fact if
the field $x$
varies rapidly compared with $\varphi $, (which can be expected as
$\varphi $ is
a background field) then the last term can be ignored and the action
becomes that of a two dimensional gravity
\begin{equation}
I'_{eff} = \frac {k}{2 \pi} \int d^2z [\overline {\partial} \varphi
\partial \varphi +
2 \varphi -4 \epsilon e^{-2 \varphi} \overline {\partial } x \partial x ],
\end{equation}
provided $x$ is rescaled to absorb $\epsilon $.

We may also boost the compact subgroup $SO(2)$ generated by $i \sigma _2$;
and study its behaviour as the boost parameter $t$
goes to infinity. The calculations are similar to the Lorentzian black
 hole
yielding ,
\begin{equation}
I_{eff} =\frac {k}{2 \pi} \int d^2z  [\overline {\partial } \varphi
\partial \varphi +
2 \varphi + 4 \epsilon e^{-2 \varphi} \overline {\partial } x
\partial x + 4 \epsilon e^{-2 \varphi}
(x^2+4)(\overline {\partial} \varphi  \partial \varphi + 1 ]
\end{equation}
with the important difference of the sign for the $x$ field terms
indicating
that again for large but finite $t$, there is a possible $c=1$
interpretation,
albeit a Euclidean one. Note that the difference of sign of the $x$ field
contribution when approaching the $E(1)$ limit naturally leads to the
expectation that
exactly at the $E(1)$ limit, the $x$ field should disappear and the one
 dimensional
Liouville field $\varphi $ remain.

The vector gauging of $SL(2,R) \ WZNW $ model Eq.(2.4) goes through the
same way with the result that for the noncompact subgroup generated by
$\sigma _3$, the
following gauge fixing conditions are allowed
\begin{equation}
u+v=0 \ \ \ \ {\rm for} \ uv<0
\end{equation}
\begin{equation}
u-v=0 \ \ \ \  {\rm for} \ uv>0
\end{equation}
and the target space metric and dilaton field become exactly of the same
form as the
axial gauge case
\begin{equation}
ds^2=-\frac {k}{2} \ \ \frac {dadb}{1-ab}
\end{equation}
\begin{equation}
\Phi = ln (1-ab) +const.
\end{equation}
with $u$ and $v$ replaced by $a$ and $b$.

For the compact subgroup generated by $i\sigma _2$. we get a similar
result when gauging the vector symmetry as the axial one,
\begin{equation}
ds^2 = \frac {k}{2} \ \  \frac {{dx'} ^2+{dy'}^2}{{x'}^2+{y'}^2 -4}
\end{equation}
\begin{equation}
\Phi =ln (x'^2+y'^2 -4)+const.
\end{equation}
where $x'=u+v, \ y' =a-b$.

The case of nilpotent subgroup $E(1)$ generated by $\sigma ^+$ is even
simpler.
The gauge fixing,
\begin{equation}
u-v=0 \ \ \ \ {\rm for} \ \ w\neq 0
\end{equation}
leaves the two gauge invariant parameters $w$ and $a+b$ to describe the
quotient manifold (for $w\neq 0$). However as in the case of the
axial gauging, the effective action becomes one dimensional, that of the
Liouville theory,
\begin{equation}
I_{eff} = \frac {k}{2\pi } \int d^2z (\partial \varphi \overline
{\partial } \varphi +2\varphi ),
\end{equation}
exactly as in (2.32), when (2.37) gauge is used and
\begin{equation}
I_{eff} = \frac {k}{2\pi} \int \frac {\partial x \overline
 {\partial } x}{4+x^2},
\end{equation}
when we use the gauge condition (2.33).
The infinite boost and the limiting procedure described above also applies
in this case .
It is remarkable that the result of axial and vector gauging become
identical, thus
satisfying duality trivially. The duality relation between the two
gaugings collapses to a selfduality for a manifold with less dimensions.
We will comment on this later in section IV.
\vskip 1cm

\section{$SL(2,R)\times U(1)$ Gauged by  $E(1)$ subgroup}
\setcounter{equation}{0}

We saw in the previous section that the target manifold of
$SL(2,R) \ WZNW$ model gauged by
its $E(1)$ subgroup , generated by the nilpotent element $\sigma ^+=
\sigma _3+ i \sigma _2$, degenerated into
a one dimensional field which we identified with the Liouville field,
thus
removing all of the geometric structure associated with the black hole
solution
obtained from gauging a non-compact subgroup $SO(1,1)$ of $SL(2,R)$.
To
gain insight into the geometric consequences of gauging a nilpotent
subgroup,
we shall therefore consider a group larger than $SL(2,R)$. The
simplest example to take is $SL(2,R) \times U(1)$ as done in Ref [4],
where a black
string was discovered . We will gauge the diagonal $E(1)\times U(1)$
subgroup rather
 than the usual
noncompact $U(1)$.

In Ref [4] the subgroup $U(1)$ generated by $\sigma _3$ was axially
gauged for the $SL(2,R) \times U(1)$ \\ $WZNW$ model. The action
\begin{equation}
\begin{array}{ll}
I(g,A)=I(g) & +\frac {k}{2 \pi } \int d^2z \ \ tr [ {\bf  {A}} \overline
{\partial } g g^{-1}+
\overline { \bf {A} } g ^{-1} \partial g + {\bf {A} }
\overline { \bf {A} } + {\bf {A}} g \overline {\bf {A} } g^{-1}]\\ \\
 & + \frac {k}{2\pi } \int d^2 z [\frac {4c}{k} (A \overline {\partial }
f +
\overline {A} \partial f)+ \frac {8c^2}{k}A \overline {A} ] +
\frac {1}{\pi } \int d^2z \partial f \overline {\partial } f
\end{array}
\end{equation}
is invariant under the transformation
\begin{equation}
\begin{array}{ll}
 & g \longrightarrow hgh \ \ \ , \ \ \ g \epsilon  SL(2,R); \ \ \ f
\longrightarrow h^c fh^c \ \ \ ,  \ \ \ f\epsilon  U(1) \\
& {\bf {A}} \longrightarrow h ({\bf {A}} + \partial )h^{-1} \ \ \ ,
\ \ \ \overline {\bf {A}} \rightarrow h^{-1}
(\overline { \bf {A} } - \overline {\partial })h
\end{array}
\end{equation}
here $h$ is an element of the subgroup $SO(1,1)$ generated by
$\sigma _3$, and $c$ is a constant specifying the charge of of the
 field $f$ under the gauge group. By $h^c$ we mean $e^{tc} $ when
 $h=e^{t\sigma_3}$ . It is worth noting that this group is nilpotent only
when it act on $SL(2,R)$ . We have used $ {\bf {A} } =A\sigma _3$, to
distinguish the Lie algebra valued gauge field $ {\bf {A} }$ from the
c-number gauge fields $A.$ Using the gauge fixing condition $a+b=0$,
(or $a-b=0$) and
integrating the gauge fields, it was found that the resultant target
manifold action
\begin{equation}
\begin{array}{ll}
I_{eff} = \frac {1}{\pi } \int d^2z [ \frac {k  }
{8 r^2 (1- \frac {\lambda }{r}) (1- \frac {1+ \lambda }{r})}\partial r
\overline {\partial } r - (1- \frac {1+\lambda }{r})
\partial \tau \overline {\partial} \tau + (1- \frac {\lambda }{r})
\partial f \overline {\partial } f & \\ \\
+ \sqrt {\frac {\lambda }{1+ \lambda}} (1- \frac { 1+\lambda }{r})
(\partial f \overline {\partial} \tau -\overline {\partial }
 f \partial
\tau)] &
\end{array}
\end{equation}
describes the  metric, where $\lambda = \frac {2c^2}{k}$
\begin{equation}
ds^2=-(1- \frac {1+\lambda }{r}) d\tau^2 + (1- \frac {\lambda }{r})
df^2+
(1- \frac {1+ \lambda }{r})^{-1}
(1- \frac {\lambda }{r})^{-1} \frac {kdr^2}{8r^2},
\end{equation}
 where $\lambda = \frac {2c^2}{k}$ and the  axion field
\begin{equation}
B_{\tau f} = \sqrt {\frac {\lambda }{1+\lambda }}
 (1- \frac {1+\lambda }{r})
\end{equation}
Here
\begin{equation}
\begin{array}{ll}
u= & e ^{\frac {\tau}{\sqrt {\frac {k}{2} (r+\lambda )}}}
\sqrt { r- (1+\lambda )}\\
v= & -e ^{-\frac {\tau}{\sqrt {\frac {k}{2} (r+\lambda )}}}
\sqrt { r- (1+\lambda )}\\
\end{array}
\end{equation}
The dilaton field is
\begin{equation}
\Phi=ln \ r+a
\end{equation}
The metric (3.4) corresponds to the black string ,
\begin{equation}
ds^2 =-(1- \frac {M}{r'})d\tau ^2 + (1-\frac {Q^2}{Mr'} )^{-1} {df}^2
 +(1- \frac {M}{r'})^{-1}(1-\frac {Q^2}{Mr'} )^{-1} \frac {kdr^2}{8r^2},
\end{equation}
where the axionic charge $Q $ is,
\begin{equation}
Q= e ^a \sqrt  {\frac {2}{k} \lambda (1+\lambda )}
\end{equation}
and mass per unit length of the black string (3.4) is
\begin{equation}
M= e^a \sqrt { \frac {2}{k}} (1+\lambda )
\end{equation}
In (3.8), $r' =e ^a \sqrt {\frac {2}{k}} r$. The black string manifold
with metric
(3.4) has
a singularity at $r=0$ and an event horizon at $r=1+ \lambda $.
 Notice that
$\lambda $ is a free parameter corresponding to the representation
of $f$ under
$SO(1,1)$.

We will now consider gauging the  $E(1)\times U(1)$ subgroup of
 $SL (2,R) \times U(1)$,
generated by $\sigma ^+= \sigma _3 + i \sigma _2$. We notice that the same
action is still invariant under the axial gauge transformation (3.2)
even if
$E(1)$ is nilpotent. Note however, that the term ${\bf {A}}
\overline { \bf {A} }$ in (3.1) is absent
now. Again, gauge fixing by setting $a+b=0$ (when $w\neq 0$),
and integrating over
the gauge field, we obtain,
\begin{equation}
\begin{array}{ll}
I_{eff} = \frac {1}{\pi} \int d^2z \left \{ \frac {w^2}{w^2 -4 \lambda }
\partial f \overline {\partial }
f + \frac {c}{w^2 -4 \lambda } [ x (\partial f \overline {\partial } w -
\partial w \overline {\partial }
f) - w ( \partial f \overline {\partial } x - \partial x \overline
{\partial }
f)] \right. & \\
 \left. + \frac {k}{ 2 w^2 (w^2 - 4 \lambda )}
 [ (w^2 - \lambda x ^2 - 4 \lambda )
\partial w \overline {\partial } w - \lambda w^2 \partial x \overline
 {\partial }
x + \lambda w x ( \partial x \overline {\partial } w + \partial w
\overline {\partial }
x)] \right  \}
\end{array}
\end{equation}
where $w=a-b-u-v$ and $x=u-v$. As this is not diagonal for the symmetric
 part, we
will make the change of variable,
\begin{equation}
y=\frac {x}{w})
\end{equation}
and end up with the final form of the effective action,
\begin{equation}
I_{eff}=\frac {1}{\pi } \int d^2z [ \frac {k}{2} \ \  \frac { \partial w
\overline {\partial }
w}{w^2} + \frac {w^2}{w^2 - 4 \lambda } ( \partial f \overline {\partial }
f - c^2\partial y \overline {\partial } y)
+ \frac  {cw^2}{w^2 -4 \lambda } (\partial f \overline {\partial } y -
\partial y \overline {\partial } f )]
\end{equation}
Note that we are working in the gauge $w \neq 0$, $a+b=0$; and therefore
the transformation (3.12) is
well defined.

 From (3.13), we find the metric and the axionic antisymmetric field,
\begin{equation}
ds^2= \frac {k}{2}  \ \ \frac {dw^2}{w^2} + \frac {w^2}{w^2 - 4 \lambda }
(df^2 -c^2dy^2)
\end{equation}
\begin{equation}
B_{fy}= - \frac {w^2}{w^2 -4 \lambda}.
\end{equation}
The dilalon field comes out as,
\begin{equation}
\Phi = ln (w^2 - 4\lambda )+ a
\end{equation}
depending only on $w$. Note that for $\lambda =0 $,  the metric
(3.14) reduces to the $SL(2,R)/ E(1)$ metric (2.21) plus a free bosonic
degree of
freedom $f$,
\begin{equation}
ds^2 \stackrel {\lambda \rightarrow 0} { \longrightarrow } \frac {k}{2}
\ \ \frac {dw^2}{w^2} +
df^2 \ \ \ \ \ \ \ {\rm as} \ \ \ \lambda \longrightarrow 0
\end{equation}
However, for $\lambda \neq 0$, the additional field $y$ remains in
 contrast to
the reduction of the degrees of freedom encountored for $SL(2,R)/E(1)$, Eq.
 (2.21).

There is also another difference between this case and the $ SL(2,R)/E(1)$.
As the dilaton field $\Phi$ (in 3.16) is not linear in the field $\varphi $
( which
is defined in $w=\pm e ^{\varphi }$), in contradistinction with
$SL(2,R)/E(1)$
case,
we can not interpret our solutions as Liouville field.
Observe also that at large distances $w\longrightarrow \infty $, the metric
(3.14) becomes
flat,
\begin{equation}
\begin{array}{ll}
ds^2 \longrightarrow &\frac {k}{2} d \varphi ^2 + df^2 - dy ^2 , \ \ \ \
{\rm as } \ w \longrightarrow \infty \\
\Phi \longrightarrow & 2\varphi +a
\end{array}
\end{equation}

The physical interpretation of our solution is that of an extremal charged
blackhole :  It can be seen from the metric (3.14) that the curvature is
\begin{equation}
R= - \frac {64 \lambda }{k} \ \  \frac {w^2 + 3 \lambda }
{(w^2 - 4 \lambda )^2}
\end{equation}
We can calculate  its mass $M$ and charge $Q$  following Ref.[4].  Notice
that the asymptotic metric is a three dimensional Lorentzian metric
$ \eta _{\mu \nu}$ which if we expand our metric (3.14) around it ,
$g_{ \mu \nu} =
\eta _{\mu \nu } + \gamma _{\mu \nu}$, and use the relation
\begin{equation}
M_{tot} = \frac {1}{2} \oint e ^{\Phi} [ \partial ^j \gamma _{ij} -
\partial _i
\gamma _{jj}+ \gamma _{ij} \partial ^j \Phi ] ds^i
\end{equation}
we obtain, for mass per unit length of the black hole
\begin{equation}
M=4 \lambda e ^a \sqrt {\frac {2}{k}}
\end{equation}

The axionic charge can be obtained from (3.15) using $ Q= \frac {1}{2}
 e^\Phi H^*$, where
$H$ is the antisymmetric third rank tensor of the axion field $B$.
The result is
\begin{equation}
Q=-4 \lambda e ^a \sqrt { \frac {2}{k}}
\end{equation}
Thus leading to an extremal black hole, $|Q|=M$.

The singularity of the black hole is at
\begin{equation}
w=\pm 2 \sqrt {\lambda },
\end{equation}
from (3.19), and the horizon can easily be seen to be at $w=0$, which is
in the
extended region of our manifold (Fig.3). Note that at $\lambda =0$ we
obtain
a flat , $R=0$, one dimensional space , as expected from the decoupling
of $SL(2,R)$
from the $U(1)$ factor. It should be noted that although we started from
arbitrary
$\lambda$, we landed with the special extremal solution $M=|Q|$ of Horne
and
Horowitz.

To understand the reason for the appearance of the special extremal
solution
we go back to the Horne and Horowitz gauged $WZNW$ model and boost the
subgroup $SO(1,1)$
used there. Thus we follow the procedure of the last section and boost
the generator $\sigma _3$
as in (2.24) and look at the limiting case of the boost parameter
 $t \longrightarrow \infty $.
However, there is a subtlety related to the transformation of the
 $U(1)$ factor under the
boost. The $U(1)$ transforms as a one dimensional representation with
weight $c$
under the $SO(1,1)$ generated by $\sigma _3$, (Eq.3.2). A priori
there is no
connection between $c$ and the weight $c_t$ of the $U(1)$ transformation
under
the $SO(1,1)$ generated by $\sigma _3 ^t$ of Eq.(2.24). However, if
the boost
is to be an isometry of the manifold, leaving the effective action (3.3)
invariant, we must choose
\begin{equation}
c_t=ce^t
\end{equation}
With this caveat we can find the limit $t\longrightarrow \infty $ of the
$SL(2,R) \times U(1)$  ,  gauged by the subgroup generated by
$\sigma _3 ^t$, and observe that the region specified by the gauge
condition $a+b=0$,
goes to our effective action (3.13) of the $SL(2,R) \times U(1) /E(1)$
theory. We also find that the singularity of the Horne and Horowitz
solution, (3.4), at $r=0$, goes to our singularity of (3.14), at
$w^2=4 \lambda $.
The horizon at $r=1 +\lambda $ going to our horizon at $w=0$. Finally the
charge and mass going to the same value
$|Q|=M= 4 \lambda \sqrt {\frac {2}{k}} e^a$
giving the expected extremal black hole solution. We can now see the
 reason for
this: the mass, charge ratio depends on $\lambda $ in the Horne and
Horwitz solution
(Eqs.(3.9) and (3.10)). But as $\lambda $ depends on the boost parameter
 $t$ and
diverges as $t \longrightarrow \infty $, the ratio $|Q|/M$ goes to unity,
giving
the extremal solution.

The region near $w=0, \ a-b=0$, can also be similarly treated.
In this gauge,the action
(3.1) becomes
\begin{equation}
I=\frac {k}{2 \pi} \int d^2z [ \frac {|\partial x|^2}{4+x^2}+ \frac {2}{k}
|\partial f|^2] + \frac {2c}{\pi} \int d^2z (A \overline {\partial} f+
\overline {A} \partial f + 2c \overline {A} A),
\end{equation}
which clearly shows the decoupling of the $U(1)$ factor from $SL(2,R)$.
Integration over the gauge field $A$ removes $f$, as it appears in the
action
as $|\partial f+ 2cA|^2$ , and the effective action becomes one dimensional
and identical to $SL(2,R)$ case in section II,
\begin{equation}
I_{eff}=\frac {k}{2 \pi} \int d^2z \frac {\partial x
\overline {\partial } x}{4+x^2}
\end{equation} .

We will next consider vector gauging the $E(1)$ subgroup of
$SL(2,R) \times U(1)$.
In the region determined by the gauge condition $u-v=0 , \ w \neq 0$,
we obtain the following effective action
\begin{equation}
I_{eff}= \frac {1}{\pi } \int d^2z [ \frac {k}{2} \ \ \frac {\partial w
\overline {\partial } w}{w^2}
+ \frac {w^2}{w^2+4 \lambda } (\partial f \overline {\partial } f -
c^2\partial y \overline {\partial } y)+
\frac {cw^2}{w^2+4 \lambda } (\partial f \overline {\partial } y -
\partial y \overline {\partial } f )]
\end{equation}
which is very similary to the axial gauge case Eq. (3.13), except
for the change of
signs. We can read off the metric and the axionic field $B$ from (3.26),
\begin{equation}
ds^2= \frac {k}{2} \ \ \frac {dw^2}{w^2} + \frac {w^2}{w^2+4 \lambda }
(df^2-c^2dy^2)
\end{equation}
\begin{equation}
B_{x y}= \frac {w^2}{w^2+4 \lambda }
\end{equation}
The dilaton field can be similarly obtained,
\begin{equation}
\Phi=ln (w^2+4 \lambda )+a
\end{equation}

In contrast to the $SL(2,R)$ case, duality between the axial and vector
gauging
is not trivial. In fact, to obtain the metric in the case of vector
 gauging
from that of the axial gauging, we must let
\begin{equation}
\lambda \longrightarrow - \lambda ,
\end{equation}
which however has to accompany a change of the sign of the $B$ field
and
consequenlly the sign of charge, in order to obtain Eq.(3.28). But
a change in sign of $\lambda = \frac {2c^2}{k}$ requires complexification
of
the representation weight $c$! The reason for the difficulty is that
in the vector gauging
$g \longrightarrow h^{-1} \ gh$, the transforma
 $f \longrightarrow h^{-c} f h^c$
leaves $f$ invariant, while our original action
\begin{equation}
\begin{array}{ll}
I(g,A)=I(g)+\frac {k}{2 \pi} \int d^2z tr [ {\bf {A} } \overline
 {\partial }
g g^{-1} - \overline {\bf {A}} g^{-1} \partial g + {\bf {A} }
\overline {\bf {A} } - {\bf {A} } g \overline {\bf {A}} g^{-1}] & \\ \\
+\frac {k}{2 \pi} \int d^2z [ \frac {4c}{k} (A \overline {\partial } f+
\overline {A} \partial f)
+ \frac {8c^2}{k} A \overline {A}] & \\
\end{array}
\end{equation}
obtained as a modification of (3.1), is invariant only with $f
\longrightarrow f+ 2 \epsilon c$.

An action invariant under $f \longrightarrow f$, has been considered by
 Ishibashi et.al. [8],
with the relevant $U(1)$ part,
\begin{equation}
I_f = \frac {1}{\pi } \int d^2z [ \partial f \overline
{\partial } f + 2ic ( \overline {A} \partial f - A \overline
 {\partial } f)]
\end{equation}

Using this action we could again consider gauging with the $E(1)$ subgroup
and
find various physical quantities such as the metric, the dilaton field,
 and the background gauge fields.
But, a clearer duality does not appear here either.
The problem is that the appropriate action
for vector gauging is Eq.(3.1), while the appropriate action for axial
gauging
is Eq.(3.33) with (3.34) replaced for the $U(1)$ part; and these are
fundamentally
different; thus duality is muddled.
\vskip 1cm

\noindent
\section{Conclusion.}
\setcounter{equation}{0}

In this work we have studied the coset models $G/H$, as conformal field
theories obtained
by gauging the subgroup $H$ of the $WZNW$ model for $G=SL(2,R)$ or
$SL(2,R) \times U(1)$,
specializing $H$ to the non semisimple subgroup $E(1)$ of $SL(2,R)$.
We found
that the two dimensional coset $SL(2,R)/ E(1)$ gives a target space
theory of only
one dimension which we identified with the Liouville field theory with
vanishing cosmological
constant, Eq.(2.23). The same type of dimensional reduction did not appear
for
$SL(2,R) \times U(1) / E(1)$, which resulted in a three dimensional
target space
theory and exhibitted on extremal black hole (or rather black string),
$|Q|=M$,
structure (Eq.3.13).

To see how the above structureis formed, we took the $SL(2,R)/U(1)$
black hole of
Witten [1], and the $SL(2,R) \times U(1) / U(1)$ of Horne and
 Horowitz [4],
and boosted the $U(1)$ gauged subgroup and studied their behaviour as
 the boost
parameter tended to infinity. As expected , at the limit of infinite boost,
 the
$SL(2,R) / U(1)$ black hole reduced to our Liouville Field of Eq.(2.23),
and
the $SL(2,R) \times U(1)$ charged black string reduced
to our
extremal charged black string of Eq.(3.13). We argued that in the $SL(2,R)$
case, there
is enlargening of the orbit of the gauge group when the boost tends to
infinity and
that region inside the black hole shrinks to a one dimensional slice in
this limit;
thus results an one dimensional theory, a phenomenon which is prevented
by
the behaviour of $\lambda $ when boosted, for the $SL(2,R) \times U(1)$
case.
In the following we will pursue the study of the course of reduction of
 the extra degree
of freedom for $SL(2,R)$ and hope to shed furthur light on this
phenomenon.

We would like to argue that the disappearance of the parameter $x$ from
the apparantly two
parameter space of the action in Eq. (2.2) , is the consequence of a
sudden
enlargement of its symmetry , as the gauge subgroup $U(1)$ tends to $E(1)$
under boost.

In fact, the gauge part of the action
\begin{equation}
I_{\rm gauge} = \frac {k}{2\pi} \int d^2z tr [{\bf A} \overline {\partial }
g g^{-1} +
\overline {{\bf A}} g^{-1} \partial g + {\bf A} \overline {{\bf A}}
+{\bf A} g \overline {{\bf A}} g^{-1} ]
\end{equation}
written in terms of $\sigma ^{\pm } = \sigma _3 \pm i \sigma _2$, takes
the form
\begin{equation}
\begin{array}{ll}
&I_{\rm gauge}  {\displaystyle = \frac {k}{2\pi} \int d^2z tr
 [A \sigma^+ \overline {\partial } g g^{-1} +
\overline {A} \sigma ^+ g^{-1} \partial g + A \overline {A}
\sigma ^+ g \sigma ^+ g ^{-1}]}\\ \\
 & {\displaystyle  + \frac {k}{2 \pi} e^{-2t} \int d^2z tr
 [ A \sigma ^ - \overline {\partial } g g^{-1} +
\overline {A} \sigma ^ -g^{-1} \partial g + 2 A \overline {A} +
 A \overline {A}
(\sigma ^ - g \sigma ^+ g^{-1} + \sigma ^+ g \sigma ^- g ^{-1})]} \\ \\
 & + O(e ^{-4 t})
\end{array}
\end{equation}
where we denote by $A$ and $\overline {A}$ the c-number gauge fields
scaled by
$e^t$. Now if we let $\overline {A} \longrightarrow -\overline {A} $
in (4.1),
this action changes into the gauge part of the action for vector gauging
(2.4),
except for $A\overline {A}$ term which has the wrong sign. But Eq.(4.2)
shows that
to leading order  in $e^{-t}$, the $A \overline {A}$ term does not appear,
and in the
infinite $t$ limit, the action has an effective vector gauge symmetry
\begin{equation}
\begin{array}{ll}
A \longrightarrow & A- h^{-1} \partial h\\
\overline {A} \longrightarrow & \overline {A}+ h^{-1} \overline {\partial}
h\\
\end{array}
\end{equation}
Although this transformation violates reality of $A$, it is formally a
symmetry of our theory
and is respected by the equations of motion and survives path integration.
 It is not difficult to check that the
vector and axial gauge transformations are compatible and that the axial
 gauge condition is respected
by the vector gauge transformation and vice versa. Then the only variable
which is both
vector and axial gauge invariant is $w=tr \sigma^+ g$, thus giving us a
one dimensional field
theory in terms of $w$.

The phenomenon of the reduction of the effective degree of freedom in
gauged $WZNW$ models,
is not restricted to our example and, we conjecture it occurs whenever the
gauged
subgroup contains nilpotent factors. There are other specific examples
[17]
and the general
question is under
investigation by the authors.
\vskip 1.2cm
\noindent
{\large Acknowledgements}

We would like to thank S.Rouhani and M. Khorrami for useful discussions.
H.A. thanks W.Nahm for his kind hospitality
at the
Physics Institute of Bonn University. H.A. also acknowledges the financial
support by Max- Planck Institute for Mathematics .

\begin{center}
{\large \sc References\\}
\end{center}
\begin{enumerate}
\item E. Witten, {\bf Phys. Rev. D44} (1991)314.; see also G.
 Mandal, A.M. Sengupta, and S.R. Wadia,
{\bf Mod. Phys. Lett. A6} (1991)1685
\item R. Dijkgraaf, H. Verlinde, and E. Verlinde, {\bf Nucl. Phys.
B371} (1992)269
\item P. Ginsparg and F. Quevedo, {\bf Nucl. Phys. B385} (1992)527
\item J. Horne and G. Horowitz, {\bf Nucl. Phys. B368} (1992)444
\item S.K. Kar and A. Kumar, {\bf Phys. Lett. B291} (1992)246
\item C.R. Nappi and E. Witten, {\bf Phys. Lett. B293} (1992)309
\item P. Horva, {\bf Phys. Lett. B278} (1992) 101
\item N. Ishibashi, M.Li, and A.R. Steif, {\bf Phys. Rev. Lett. 67}
(1991)3336
\item D. Gershon, {\bf Exact solutions of four dimensional Black Holes
in string Theory,}
TAUP-1939-91, Dec 1991
\item I. Bars and D. Nemechansky, {\bf Nucl. Phys. B348} (1991)89;
I. Bars, Phys. Lett. B293 (1992)315
\item A. Sen, {\bf Rotaing Black Hole Solution in Heterotic String
Theory},
TIFR / TH / 92-20
\item E.B. Kiritsis, {\bf Mod. Phys. Lett. A6} (1991) 2871
\item L. Feh\'er, L. O'Raifeartaigh, P. Ruelle, I. Tsutsui, and A.
Wipf, {\bf Phys. Rep. 222} (1992) 1
\item A. Alekseev and S. Shatashvili, {\bf Nucl. Phys. B323} (1989) 719
\item M. Bershadsky and H. Ooguri, {\bf Com. Math. Phys. 126} (1989) 49
\item F. Ardalan, {\bf 2D Black Holes and 2D Gravity}, Proc. of Workshop
on
Low Dimensional Topology and Quantum Field Theory, Isaac Newton Inst.,
Sept. 1992, to be
published by Cambridge Press
\item F.Ardalan and A.Ghezelbash, in Preparation
\end{enumerate}
\end{document}